%% file: main.tex
\documentclass[conference]{IEEEtran}
\usepackage{subcaption}
\usepackage{etoolbox}
\usepackage{xcolor}
\usepackage{hyperref}
\usepackage{amssymb}
\usepackage{xurl}

\ifCLASSINFOpdf
  \usepackage[pdftex]{graphicx}
\else

\fi

\begin{document}
\title{Trust or Bust: A Survey of Threats in \\ Decentralized Wireless Networks}

\author{\IEEEauthorblockN{Hetvi Shastri\IEEEauthorrefmark{1}, Akanksha Atrey\IEEEauthorrefmark{2}, Andre Beck\IEEEauthorrefmark{2}, Nirupama Ravi\IEEEauthorrefmark{2}}
	\IEEEauthorblockA{\IEEEauthorrefmark{1}University of Massachusetts Amherst, \IEEEauthorrefmark{2}Nokia Bell Labs\\
		hshastri@cs.umass.edu, \{akanksha.atrey, andre.beck, nirupama.ravi\}@nokia-bell-labs.com}
 }

\pagestyle{plain}

\IEEEoverridecommandlockouts
\makeatletter\def\@IEEEpubidpullup{6.5\baselineskip}\makeatother
\IEEEpubid{\parbox{\columnwidth}{
		Workshop on Security and Privacy of Next-Generation Networks \\ (FutureG) 2025 \\
		24 February 2025, San Diego, CA, USA \\
		ISBN 979-8-9919276-7-3 \\ 
		https://dx.doi.org/10.14722/futureg.2025.23046 \\   
		www.ndss-symposium.org
}
\hspace{\columnsep}\makebox[\columnwidth]{}}

\maketitle

\input{abstract}

\IEEEpeerreviewmaketitle

\input{introduction}

\input{motivation}

\input{attacks}

\input{countermeasures}

\input{conclusion}

\bibliographystyle{IEEEtran}

\bibliography{ref}

\appendix
\input{appendix}

\end{document}

%% file: abstract.tex
\begin{abstract}
    The recent emergence of decentralized wireless networks empowers individual entities to own, operate, and offer subscriptionless connectivity services in exchange for monetary compensation. While traditional connectivity providers have built trust over decades through widespread adoption, established practices, and regulation, entities in a decentralized wireless network, lacking this foundation, may be incentivized to exploit the service for their own advantage. For example, a dishonest hotspot operator can intentionally violate the agreed upon connection terms in an attempt to increase their profits. In this paper, we examine and develop a taxonomy of adversarial behavior patterns in decentralized wireless networks. Our case study finds that provider-driven attacks can potentially more than triple provider earnings. We conclude the paper with a discussion on the critical need to develop novel techniques to detect and mitigate adversarial behavior in decentralized wireless networks.
\end{abstract}

%% file: introduction.tex
\section{Introduction}
\begin{figure}[h]
    \centering
    \includegraphics[width=\linewidth]{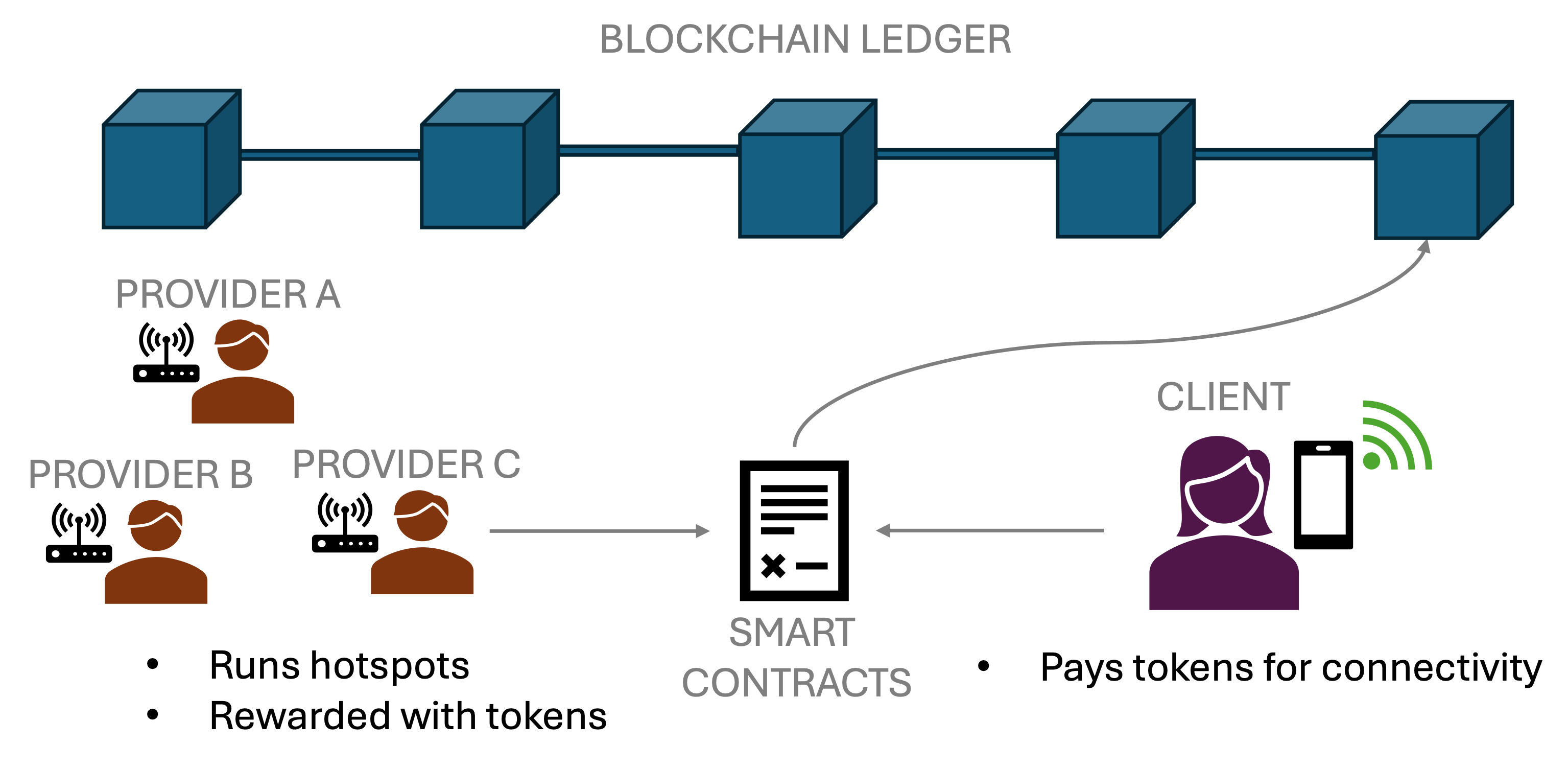}
    \caption{A simplified view of DeWi networks.}
    \label{fig:denets-setup}
\end{figure}

Traditional wireless network connectivity, provided by a few centralized mobile network operators (MNOs), relies heavily on strict access control, subscription-based service models, and operator-controlled usage accounting. The emergence of decentralized wireless (DeWi) networks in recent years has shown great potential to disrupt the traditional connectivity market and create new opportunities for both operators and users. Examples of DeWi networks currently being built include Helium Mobile/IoT \cite{helium}, XNET \cite{xnet}, World Mobile \cite{worldMobile}, and WayRu \cite{wayru}. The infrastructure making up DeWi networks is owned, operated, and deployed by a distributed group of participants rather than a single centralized entity. This decentralized model lowers the barrier to entry and thus allows for more competition and a faster crowd-sourced network build out. It also gives users more control over which network they use for connectivity. Blockchains and smart contracts are the building blocks of decentralized connectivity--replacing legal agreements commonly used in traditional networks and providing payment guarantees. Figure \ref{fig:denets-setup} shows a high-level view of DeWi networks.

Traditional connectivity providers have built trust over decades through widespread adoption, established practices, and regulatory oversight. This foundation ensures a level of reliability and security that users have come to expect. In contrast, the potentially large number of hotspot operators in a decentralized wireless network leads to a lack of trust, making it more likely that some may act maliciously \cite{evilTwin, routerAttack, pocGaming}. Furthermore, the decentralized nature of such systems opens up opportunities for users themselves to exploit and cheat the system, potentially compromising the integrity and security of the service. For example, a semi-malicious hotspot operator can intentionally violate the agreed upon connection terms by actively limiting the bandwidth for a subset of users or admit more users than they can support in an attempt to increase their profits. On the other hand, a malicious user may degrade the operator's ability to provide their service through forced retransmissions of intentionally dropped packets.

As the connectivity landscape evolves, it is crucial to develop new frameworks and safeguards to address the unique security challenges posed by DeWi networks. In this paper, we investigate these unique challenges and develop a taxonomy of adversarial behavior patterns in DeWi networks. Our contributions are summarized as follows:

\begin{enumerate}
    \item We present a case study of provider-driven attacks in a DeWi network and demonstrate the ability of providers to more than triple their profits with minimal effort.
    \item We develop a taxonomy of adversarial behavior in DeWi networks from the provider, user, and third-party points of view. The taxonomy differentiates between confidentiality attacks, integrity attacks, and capacity attacks.
    \item We discuss countermeasures and highlight the need for novel solutions to mitigate attacks in DeWi networks.
\end{enumerate}

%% file: motivation.tex
\section{Background and Motivation}
\label{sec:motivation}
DeWi networks allow untrusted individuals or small businesses to provide users with a diverse set of connectivity options such as Wi-Fi, LoRaWAN, LTE, 5G, or next generation networks. The network infrastructure such as hotspots, gateways, or small cells are owned and operated by individuals or organizations rather than a central authority.

While centralized networks typically rely on legal agreements between users and providers for building trust, they typically do not provide open roaming capabilities to other networks or highly granular payment models. Users are often locked into a single service provider, preventing them from accessing the best available connectivity option at any given time. 
For example, a user may want to have a high speed connection for a short duration at a particular location and be charged on a pay per use basis. DeWi networks offset these limitations yet lack the established trust and reputation of incumbent providers given the potentially large number of players involved. This leads to potential threats. 

To demonstrate the feasibility and threats of DeWi networks, consider Alice who works in a major metropolitan city. While on her way to work, she is used to facing a disruption in network connectivity (e.g., due to overload on the network infrastructure). To complete an important task, Alice opts to join a DeWi hotspot which follows a pay-as-you-go model based on the total connection time.  

While the above model offers interesting opportunities for next-generation networks, it also comes with threats that traditional networks do not encompass. In the above setup, a malicious hotspot provider can intentionally degrade the network quality to earn higher profits by increasing Alice's connection time to the hotspot. We perform a case study on an experimental DeWi network to examine this facet. The setup details can be found in Appendix \ref{appendix:experimental-setup}.

\begin{figure}[ht]
    \centering
    \includegraphics[width=0.35\textwidth]{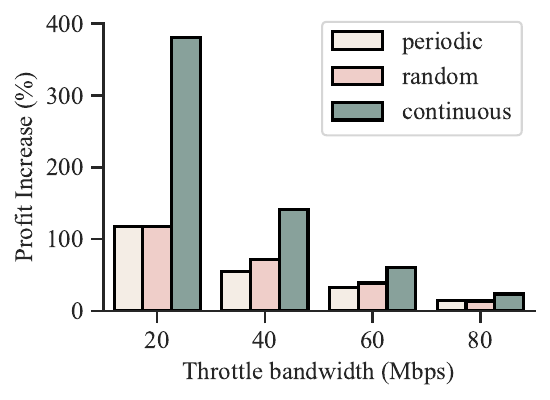}
    \caption{Effect of reducing download bandwidth from 100Mbps on the provider's monetary benefits.}
    \label{fig:experiment_1}
\end{figure}

~\autoref{fig:experiment_1} demonstrates the effects of degrading the bandwidth on the amount of time it takes to complete a file download. To avoid the risk of getting caught, the malicious provider can also throttle the bandwidth periodically (e.g., 10 seconds every 5 seconds) or randomly (e.g., once for 10 seconds in the entire session). As shown in ~\autoref{fig:experiment_1}, throttling the user's bandwidth for an entire session allows the provider to achieve exponential gains up to 378\% whereas throttling it periodically or randomly leads to a linear increase in earnings. Appendix \ref{appendix:evaluation} includes similar findings for upload and streaming tasks. This highlights the ability of the malicious hotspot provider to achieve higher gains with more risk.

\begin{figure}[ht]
    \centering
    \includegraphics[width=0.35\textwidth]{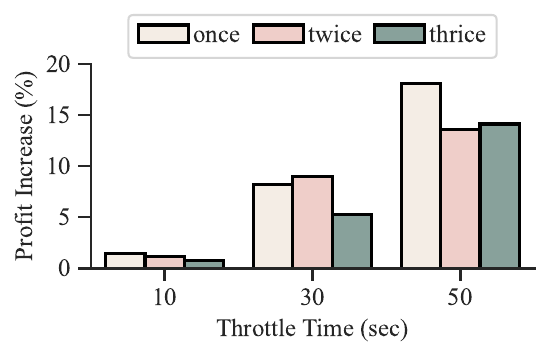}
    \caption{Effects of throttling bandwidth from 100Mbps to 50Mbps in random intervals of one, two, and three chunks.}
    \label{fig:experiment_2}
\end{figure}

The hotspot provider can also smartly throttle the bandwidth to fool the user. Consider a situation where a connected user performs a file download, streams a video, and then uploads a file during a session. ~\autoref{fig:experiment_2} demonstrates the effect of the hotspot provider throttling bandwidth in multiple intervals. For example, a provider can continuously throttle for 30 seconds, two 15 seconds chunks, or three 10 second chunks. By splitting the throttling intervals, a malicious provider can decrease their chance of getting caught while achieving the same effects in their profits. This highlights the ability of the adversary to conduct multiple short periods of bandwidth throttling while ensuring the same impact.

%% file: attacks.tex
\section{Attack Taxonomy}
\label{sec:attacks}
Wireless networks, spanning WLANs, cellular networks, Bluetooth, and Zigbee, are generally susceptible to various security threats due to their open communication channels.
Security attacks can be classified by the attacker's location (external versus internal) \cite{Malik2023} and by the intrusiveness of the attack (passive versus active) \cite{4625802}. Other works have focused on categorizing attacks in the mobile networking space based on their target layers, such as eavesdropping attack for physical layer, denial-of-service (DoS) attack for network layer, and malware attack for application layer~\cite{7547270,Korolkov2021AnalysisOA}. Some surveys provide a comprehensive understanding and taxonomy of different attacks, such as jamming attacks~\cite{5343062,9733393,shi2022launch} and rogue attacks~\cite{10.1007/s11277-016-3390-x,8770554,article}.

While users are not immune to threats from traditional connectivity providers \cite{go2013towards}, these threats are exacerbated in DeWi networks due to the lack of trust, anonymity of the involved parties, and finality of blockchain transactions. Without a trusted central entity, providers and users are in the difficult position of having to detect and mitigate adversarial behavior themselves.

\begin{figure*}[ht]
    \centering
    \includegraphics[width=\textwidth]{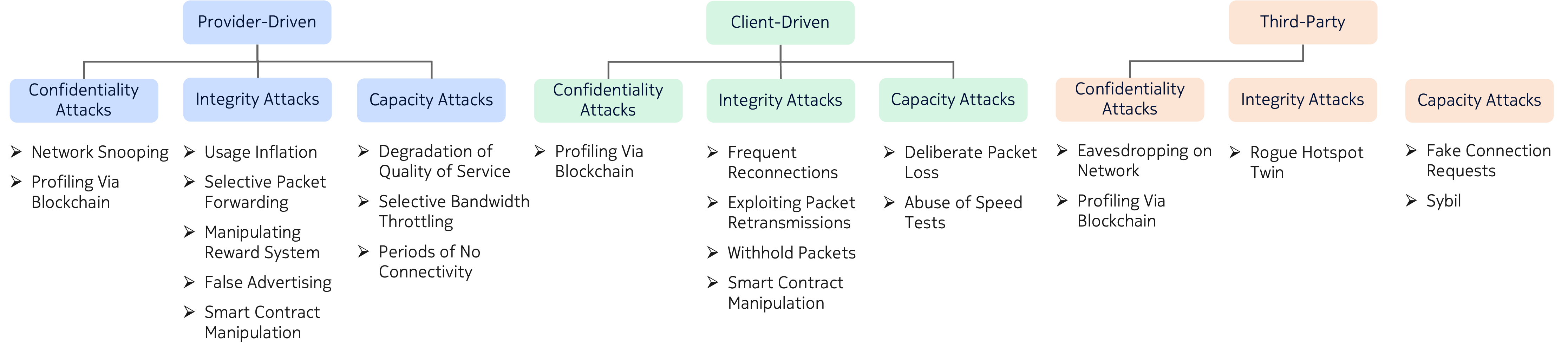}
    \caption{Taxonomy of adversarial attacks in decentralized wireless networks.} 
    \label{fig:taxonomy}
\end{figure*}

In this section, we present a comprehensive taxonomy of adversarial attacks in DeWi networks, categorized into provider-driven, client-driven, and third party attacks.
These attacks are primarily motivated by DeWi networks' pay-as-you-go model, which incentivizes providers, clients, or third parties to engage in malicious activities for financial gain as demonstrated in Section \ref{sec:motivation}. Each attack is further classified into confidentiality, integrity, or capacity attack based on its impact on the DeWi network. Confidentiality attacks allow unauthorized data access, integrity attacks allow inconsistent and improper modifications to data measurements and service offerings, and capacity attacks affect the provider's service delivery or the client's service utilization. Figure \ref{fig:taxonomy} summarizes these attacks.

\subsection{Provider-Driven Attacks}
We first outline attacks against users committed by malicious connectivity providers. 

\subsubsection{Confidentiality Attacks}
Privacy concerns in DeWi networks arise due to the fact that all user traffic passes through infrastructure owned and operated by various small footprint connectivity providers versus a single trusted entity. As such, a malicious provider is in a position to intercept and analyze user traffic which may potentially contain sensitive data. Even when transport-layer security protocols are used to encrypt application-layer connections, such as HTTP sessions, a provider could still build a detailed user profile based on the sites they communicate with and use that information for targeted scam/phishing attacks. Furthermore, DeWi networks offer transparency and immutability via blockchains, enabling user profiling based on blockchain transactions, irrespective of their connection status to the provider's hotspot. This allows anyone in the network to access historical user transactions. While blockchains enhance transparency, the inherent anonymity in DeWi networks also makes it challenging to hold individuals accountable for malicious activities.

\subsubsection{Integrity Attacks}
The most common billing model in DeWi networks is pay-as-you-go where users are charged for their actual data usage or connection time according to the agreed upon connection terms. A malicious provider could attempt to gain a monetary benefit by charging the user surreptitiously for an inflated amount of data usage (e.g., by accounting for network-induced packet retransmissions \cite{go2013towards}). Providers could also compromise the promised connectivity by filtering or censoring the user's online activity via selective forwarding \cite{1203362}. DeWi networks offer incentives that can be manipulated for higher rewards. For example, if providers are rewarded based on their bandwidth, they can artificially inflate their bandwidth to receive higher rewards via rushing packets \cite{sheng2022proof}. Alternatively, if the reward system is based on their ability to prove coverage in a certain location, they can conduct alternate reality attacks by spoofing GPS coordinates and radio frequency signals \cite{haleem2018decentralized}.  Finally, DeWi networks that allow hotspot advertising are susceptible to false advertising, highlighting the need for a verification mechanism.

\subsubsection{Capacity Attacks}
If the agreement between the user and the connectivity provider also includes quality of service (QoS) parameters, such as minimum bandwidth, the provider could intentionally violate the QoS terms which would allow them to admit more clients then they could otherwise serve by selectively withholding packets \cite{sheng2022proof} or force users to remain connected for longer periods of time. A more sophisticated malicious provider could throttle a user's bandwidth or violate other QoS parameters selectively and temporarily to avoid detection as shown in Section \ref{sec:motivation}. A more brazen attack may consist of periods of no connectivity at all while the user is still paying for it.
\subsection{Client-Driven Attacks}
Client-driven attacks consider users in a DeWi network who attempt to fool the system for monetary benefits.

\subsubsection{Confidentiality Attacks}
While the use of blockchains in DeWi networks enhances transparency, enabling users to making informed decisions on cost effectiveness and network capabilities, they also reveal usage patterns of other users, posing a significant privacy risk. A well-meaning user might unintentionally profile others when analyzing blockchain transactions.

\subsubsection{Integrity Attacks}
Decentralized networks provide on-demand services to users, which can be exploited by attackers seeking financial gain due to the network's flexible and dynamic nature. Frequent reconnections present a credible threat, allowing users to avoid charges, exploit promotions such as free data or reduced rates for new connections, bypass data usage or connection time limits, and/or game the system by artificially inflating activity and gain more rewards. Users can also exploit bugs in the DeWi's smart contract code for reward manipulation and token theft, abuse packet retransmissions for free-ride tunneling \cite{go2013towards}, and engage in packet withholding to fake QoS violations.
All of these attacks are more prominent without a centralized security layer. 

\subsubsection{Capacity Attacks}
Users can attempt to execute a smaller-scale DoS attack by falsely claiming they did not receive the intended service via intentional packet loss. While the user's intention would remain to get a refund or avoid paying for the service, deliberate packet loss will lead to the network retransmitting the packets, increasing communication, and reducing the network's efficiency by exhausting the resources \cite{1039518}.
Similarly, the abuse of free services such as speed tests could also result in network overload. Due to the smaller scale of the connectivity providers in DeWi networks, even a single user can have a significant impact with these attacks.

\subsection{Third Party Attacks}
Finally, we examine attacks that can be performed by third party.

\subsubsection{Confidentiality Attacks}
A malicious third party could exploit the openness of DeWi networks to eavesdrop on network communication via wormholes \cite{1420257}, thereby gaining access to sensitive information. Such activities may reveal users behaviors, usage patterns, or provider operations, potentially compromising privacy and security. Similarly, the third party (e.g., a competitor) can also snoop on blockchain transactions to profile users.

\subsubsection{Integrity Attacks}
While DeWi networks offer flexibility and exclusivity by allowing anyone to become a provider, they are still vulnerable to misuse through unauthorized twin or rogue hotspots. By imitating trusted providers, these twins can confuse users, disrupt fair pricing, and undermine trust in the network. Small scale connectivity providers will need more sophisticated tools to protect themselves against rogue twins. 
\subsubsection{Capacity Attacks}
Similar to traditional networks, third-party attackers can also execute various common attacks, including denial of service attacks \cite{4431860}, where a malicious actor floods a hotspot with excessive fake connection requests, overwhelming its capacity to handle legitimate requests. The decentralized and distributed nature of connectivity further makes these attacks more plausible. This overload can result in significant service disruption, causing the hotspot to become slow or entirely unavailable to legitimate users.
Similarly, third-party attackers may also launch Sybil attacks, wherein the attacker operates as an unauthorized external entity that generates multiple fake identities to manipulate the decentralized system~\cite{haleem2018decentralized}. The attacker can attempt to interfere with the consensus mechanism of the blockchain, thereby decreasing the capacity of honest providers. Computationally expensive consensus mechanisms such as Proof-of-Work make it nearly impossible for Sybil attacks in practice.

%% file: countermeasures.tex
\section{Countermeasures}
The lack of trust among the participants in DeWi networks introduces challenges in protecting against malicious behavior. Existing countermeasures, such as virtual private networks (VPN), quantitative evaluation of eavesdropping probability using cyber physical features~\cite{7982235,wu2020blueshield}, and novel encoding methods~\cite{10695123} suffice to protect against confidentiality attacks. However, existing solutions do not directly apply towards integrity and capacity attacks in DeWi networks.

Prior works have developed methods for estimating and monitoring throughput via active and passive means \cite{molavi2015identifying, balasingam2019detecting, lee2020perceive, sheng2022proof, wu2022lossdetection}. Such methods can be integrated in decentralized networks to establish integrity. One such work enables multiple clients to actively test and collaborate in measuring the bandwidth of a wireless access point's backhaul link \cite{sheng2022proof}. Another work considers relaying and replaying the traffic behind a VPN to test application wise throttling by Internet service providers \cite{molavi2015identifying}. These works are limited by their inability to process data in real-time and distinguish between naturally occurring and intentional QoS degradation. In real-time scenarios with intelligent bandwidth throttling intervals, identifying the optimal moment for active testing becomes challenging. Bauer has also investigated the variabilities in broadband speed tests \cite{bauer2010understanding}, highlighting the need to verify network performance robustly in real time using actual network traffic data. Meanwhile, there is limited work that aims at protecting providers against user-driven attacks. Future research should prioritize the development of trustless systems capable of mutual authentication, access management, and real-time performance verification, while also addressing the challenges of resource constraints on user devices and hotspot hardware.

Few works have attempted to detect adversarial behavior in DeWi networks. Anand et al.'s incremental service level agreements to compare provider and client bitrate measurements \cite{anand2022trust} are vulnerable to provider-driven bandwidth throttling. This approach also creates a single point of failure by requiring clients to trust providers for accurate per-user bitrate accounting. Sheng et al. have proposed a method to securely verify the physical location of a device in a decentralized network using internet delays \cite{sheng2024bft}. However, this does not prevent smart attackers from tampering with Internet delays. Finally, recent works have proposed a system that rewards external nodes for continuously monitoring and verifying transactions in a decentralized network \cite{west2023dnextg, sheng2024bft}. These are resource intensive solutions which can introduce complexities in maintaining the system's security and reliability.

%% file: conclusion.tex
\section{Conclusion}
In this paper, we examined the implications of adversarial attacks in decentralized wireless networks. Our case study of provider-driven attacks demonstrated that providers can earn up to 378\% more profit via selective bandwidth throttling attacks. We then presented a comprehensive taxonomy of attacks in DeWi networks from provider, user, and third-party perspectives covering confidentiality, integrity, and capacity of the network. Finally, we presented countermeasures to protect against adversarial behavior in networks. We acknowledge and highlight the need for future research to explore more intricate attack scenarios as well as build robust and generalizable countermeasures for adversarial behavior in DeWi networks.

%% file: appendix.tex
\subsection{Case Study}
We perform a case study to evaluate the effects of provider-driven network quality degradation on the monetary benefits for the provider.

\begin{figure}[ht]
    \centering
    \includegraphics[width=0.35\textwidth]{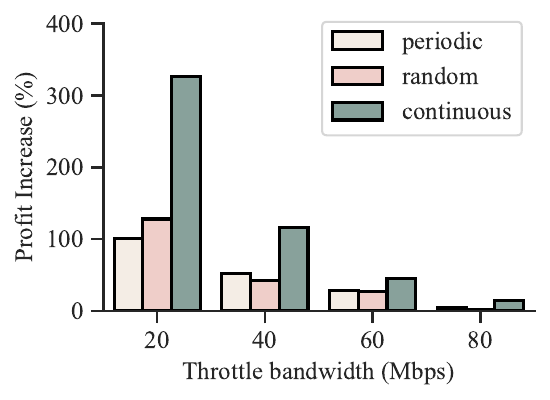}
    \caption{Effect of reducing upload bandwidth from 100Mbps on the provider's monetary benefits.}
    \label{fig:experiment_1-upload}
\end{figure}

\begin{figure}[ht]
    \centering
    \includegraphics[width=0.35\textwidth]{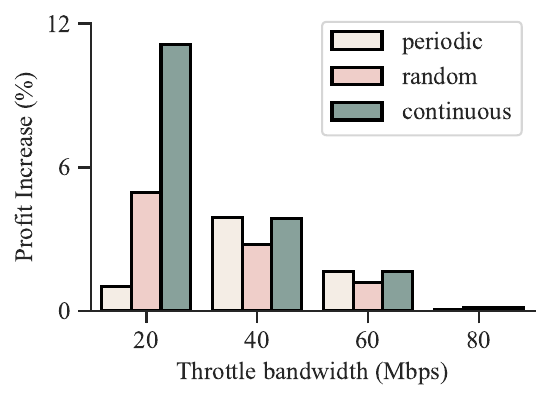}
    \caption{Effect of reducing stream bandwidth from 100Mbps on the provider's monetary benefits.}
    \label{fig:experiment_1-stream}
\end{figure}

\subsubsection{Experimental Setup}
\label{appendix:experimental-setup}

The experimental setup consists of three primary entities: the hotspot provider, the content server, and the client. The hotspot provider was setup up as a TP link router connected via an access point deployed on a Raspberry Pi (RPi) 4 model B with a Broadcom BCM2711, quad-core Cortex-A72 (ARM v8) 64-bit
SoC @ 1.5GHz, and a Linux operating system. We used the TC tool~\cite{stanic2001tc} to manage network traffic and the \texttt{netem} feature to simulate real-world network conditions like limited bandwidth. The client is deployed on a Windows machine with Intel Core i7 processor. The client connects to this Wi-Fi network and performs three tasks: downloading a 500 MB file, uploading a 100 MB file, and streaming a video. For video streaming, we used dash.js JavaScript Reference Client \cite{dashjs}, an open-source DASH video player, with a 1-minute 4K version of Big Buck Bunny~\cite{10.1145/1504271.1504321} as the test video. All files and video content were hosted on the content server, which ran an Nginx web server to handle the tasks. All experimental results are averaged over ten iterations.

\subsubsection{Evaluation}
\label{appendix:evaluation}

~\autoref{fig:experiment_1-upload} demonstrates the effects of throttled bandwidth on the provider's profit for a upload task. To avoid the risk of getting caught, the malicious provider can also throttle the bandwidth periodically (e.g., 10 seconds every 5 seconds) or randomly (e.g., once for 10 seconds in the entire session). The profit patterns are similar to the download task shown in ~\autoref{fig:experiment_1}, easily achieving up to 328\% more profit with selective degradation of QoS.

~\autoref{fig:experiment_1-stream} demonstrates the effects of throttled bandwidth on the provider's profit for a video streaming task. While the trends between the continuous, periodic, and random attacks are consistent with the download and upload tasks, the profit gain is much less. We attribute this to the fact that streaming videos requires loading and buffering the video in advance for enhanced user experience. Hence, unless the degradation occurs when the video is being loaded or buffered, the impact of the throttling will not make as much of a difference on the overall time it takes to watch the video.